\documentclass{egpubl}
\usepackage{hpg25DL}

\WsPaper         % uncomment for final version of EG Workshop contribution
 \electronicVersion % can be used both for the printed and electronic version

\ifpdf \usepackage[pdftex]{graphicx} \pdfcompresslevel=9
\else \usepackage[dvips]{graphicx} \fi

\PrintedOrElectronic

\usepackage{t1enc,dfadobe}

\usepackage{egweblnk}
\usepackage{cite}
\usepackage{pgfplots}
\usepackage{tikz}
\usetikzlibrary{math, calc, positioning, spy, backgrounds}
\usepackage{contour}
\usepackage{multirow}
\usepackage{amsmath}
\usepackage[nice]{nicefrac}
\usepackage{algorithm2e,listings}
\usepackage{breakurl}

\title[HW NBTC]%
      {Hardware Accelerated Neural Block Texture Compression \\ with Cooperative Vectors}

\makeatletter
\renewcommand*{\@fnsymbol}[1]{*}
\makeatother

\author[L. Belcour \& A. Benyoub]
{
  \parbox{\textwidth}{\centering L. Belcour\thanks{Authors share equal contribution to this work}$^{,1}$ and A. Benyoub$^{*,1}$} \\
  {\parbox{\textwidth}{\centering $^1$Intel Labs, France}}
}
\begin{document}

\newcommand{\change}[1]{
  { #1 }
}
\newcommand{\numlayers}{L}
\newcommand{\numtextures}{N}
\newcommand{\mlpdepth}{D}
\newcommand{\variantA}{\textsc{VarA}}
\newcommand{\variantB}{\textsc{VarB}}
\newcommand{\uv}{\mathbf{uv}}
\newcommand{\xvec}{\mathbf{x}}
\newcommand{\cvec}{\mathbf{c}}
\newcommand{\fvec}{\mathbf{f}}

\teaser{
  \vspace{-20pt}
  \centering
  \setlength{\fboxsep}{0pt}%
  \begin{tikzpicture}[]
    \draw[color=white] (0, 0) node (A) {} rectangle (0, 4cm) node (B) {};
    \node[above=20pt of A.north west, anchor=south west, yshift=-10pt] (latent) { \fbox{\includegraphics[width=2cm]{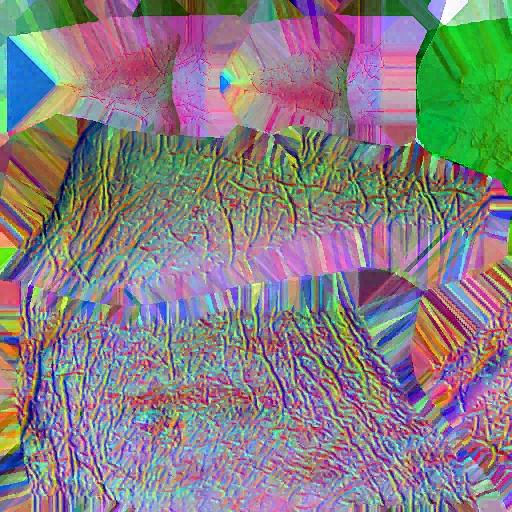}} };
    \node[right=0pt of latent.north west, anchor=north west, yshift=8pt, xshift=12pt] (latent1) { \fbox{\includegraphics[width=1.8cm]{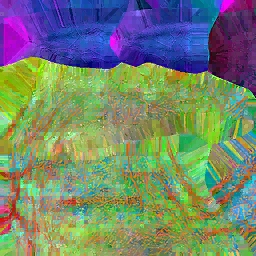}} };
    \node[right=0pt of latent1.north west, anchor=north west, yshift=8pt, xshift=12pt] (latent2) { \fbox{\includegraphics[width=1.6cm]{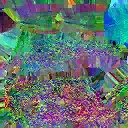}} };
    \node[right=0pt of latent2.north west, anchor=north west, yshift=8pt, xshift=14pt] (latent3) { \fbox{\includegraphics[width=1.5cm]{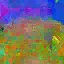}} };
    \node[above=2pt of latent.south]  { \contour{black}{\color{white}\tiny{ $W_0 \times H_0$}} };
    \node[above=2pt of latent1.south] { \contour{black}{\color{white}\tiny{ $W_1 \times H_1$}} };
    \node[above=2pt of latent2.south] { \contour{black}{\color{white}\tiny{ $W_2 \times H_2$}} };
    \node[above=2pt of latent3.south] { \contour{black}{\color{white}\tiny{ $W_3 \times H_3$}} };

    \node[right=95pt of latent, yshift=0pt,] (output_diff) { \fbox{\includegraphics[width=2cm]{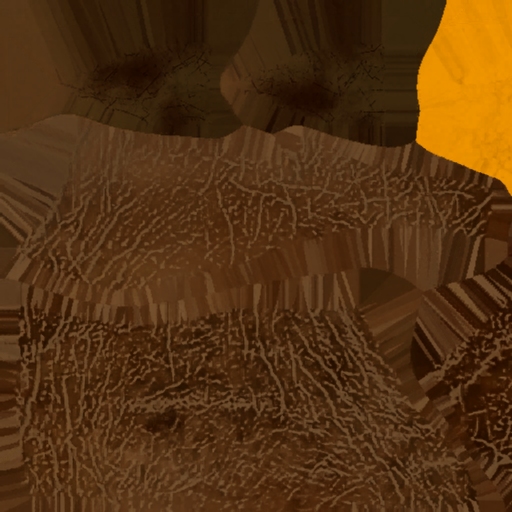}} };
    \node[right=0pt of output_diff.north west, anchor=north west, yshift=8pt, xshift=8pt] (output_disp) { \fbox{\includegraphics[width=2cm]{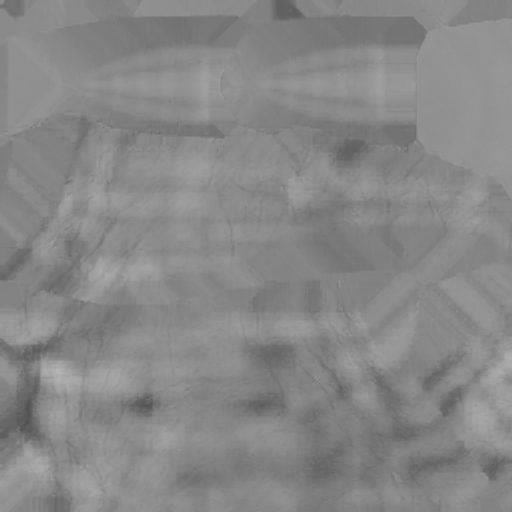}} };
    \node[right=0pt of output_disp.north west, anchor=north west, yshift=8pt, xshift=8pt] (output_mask) { \fbox{\includegraphics[width=2cm]{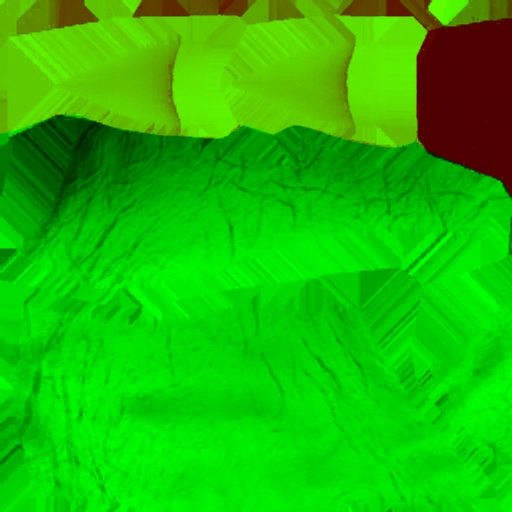}} };
    \node[right=0pt of output_mask.north west, anchor=north west, yshift=8pt, xshift=8pt] (output_metal) { \fbox{\includegraphics[width=2cm]{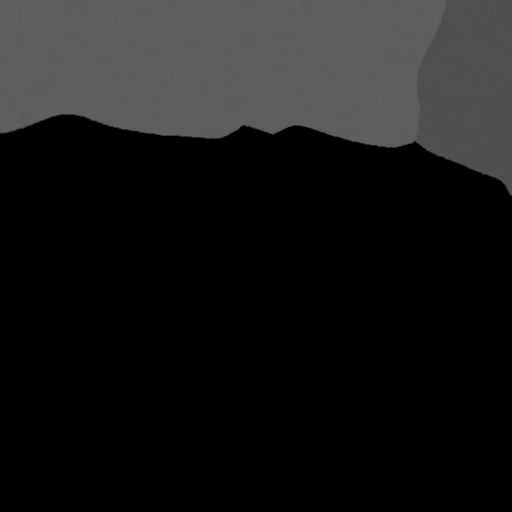}} };

    % \draw (0.5\linewidth, 0) node (A) {} rectangle (\linewidth, 4cm) node (hello) {};
    \begin{scope}
    \clip (0.5\linewidth, 0) rectangle (\linewidth, 4cm);
    \node[right=0.5\linewidth of A.north west, anchor=south west, yshift=-20pt] (output) { {\includegraphics[width=0.5\linewidth]{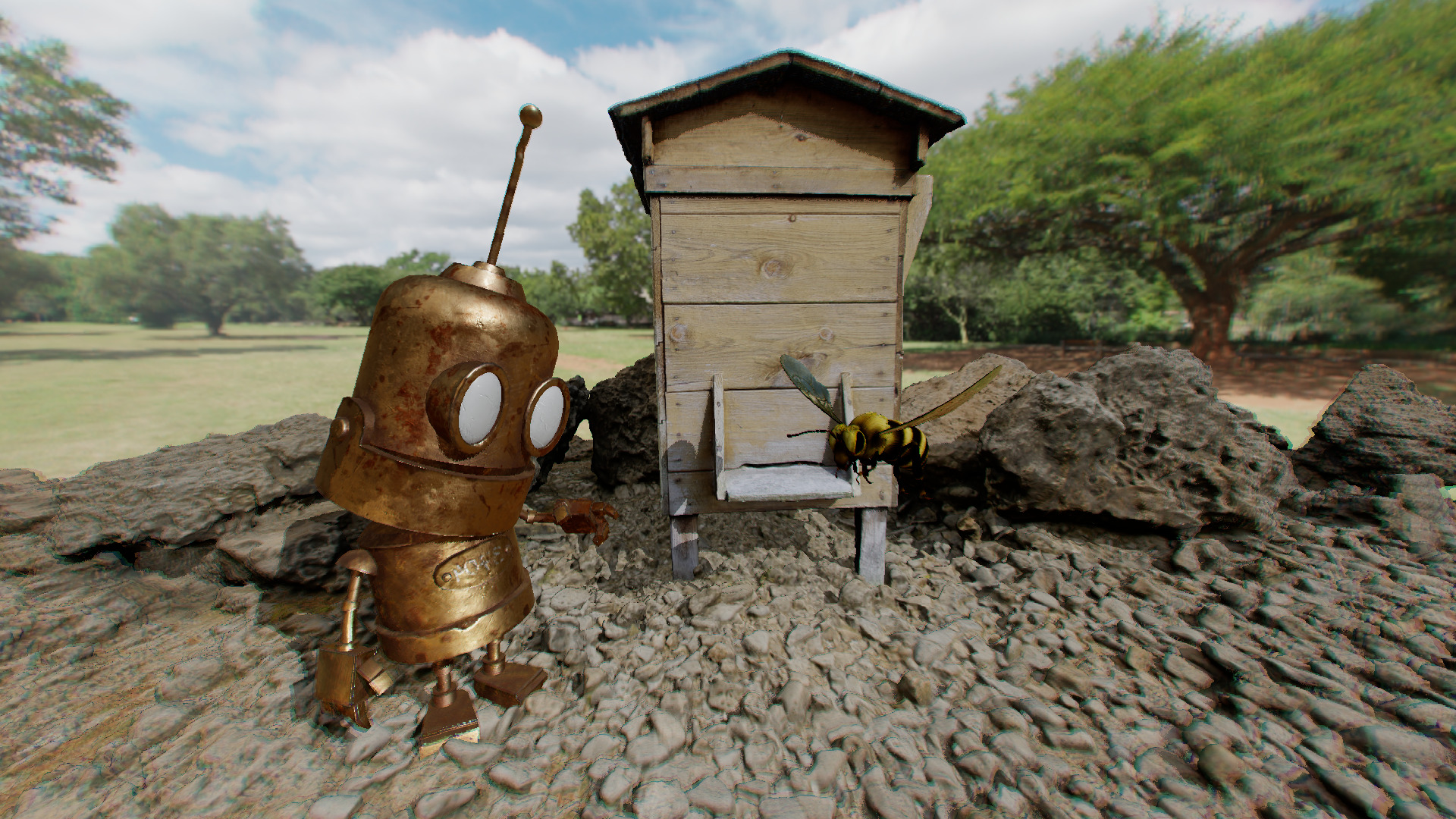}} };
    \end{scope}
    \draw (0.5\linewidth, 0) rectangle (\linewidth, 4cm);

    \node[right=0.0\linewidth of B.north, yshift=5pt, align=center, text width=0.5\linewidth] {\textbf{network architecture}};
    \node[right=0.5\linewidth of B.north, yshift=5pt, align=center, text width=0.5\linewidth] {\textbf{rendering} - $13$ \textit{ms (software FMA)} / $0.55$ \textit{ms (cooperative vectors)} };

    \begin{scope}[xshift=3.2cm, yshift=3.5cm]
        \begin{scope}

\newcommand{\inputnum}{2} 
\newcommand{\hiddennum}{3}  
\newcommand{\outputnum}{2} 

% Input Layer
\foreach \i in {1,...,\inputnum} {
	\node[draw, thick, circle, minimum size=4mm, fill=orange!30] (Input-\i) at (0,-\i) {};
}

% Hidden Layer
\foreach \i in {1,...,\hiddennum} {
	\node[draw, thick,  circle, minimum size=4mm, fill=teal!50, yshift=(\hiddennum-\inputnum)*5 mm] (Hidden-\i) at (1,-\i) {};
}

% Output Layer
\foreach \i in {1,...,\outputnum} {
	\node[draw, thick,  circle, minimum size=4mm, fill=purple!50, yshift=(\outputnum-\inputnum)*5 mm] (Output-\i) at (2,-\i) {};
}

% Connect neurons In-Hidden
\foreach \i in {1,...,\inputnum} {
	\foreach \j in {1,...,\hiddennum} {
		\draw[->, shorten >=1pt] (Input-\i) -- (Hidden-\j);	
	}
}

% Connect neurons Hidden-Out
\foreach \i in {1,...,\hiddennum} {
	\foreach \j in {1,...,\outputnum} {
		\draw[->, shorten >=1pt] (Hidden-\i) -- (Output-\j);
	}
}

% Inputs
% \foreach \i in {1,...,\inputnum} {
% 	\draw[<-, shorten <=1pt] (Input-\i) -- ++(-1,0);
% }

% Outputs
% \foreach \i in {1,...,\outputnum} {
% 	\draw[->, shorten <=1pt] (Output-\i) -- ++(1,0);
% }

\end{scope}
    \end{scope}
    \node[] at (0.07\linewidth, 0.25cm) { \footnotesize{ \textit{ BC1 textures }} };
    \node[] at (0.23\linewidth, 0.22cm) { \footnotesize{ \textit{ Multi-Layer Perceptron }} };
    \node[] at (0.38\linewidth, 0.25cm) { \footnotesize{ \textit{ PBR textures }} };
  \end{tikzpicture}
  \vspace{-17pt}
  \caption{
    \textbf{Rendering physically-based materials using neural block texture compression.} We compress PBR texture sets using block compressed latent images decoded by a multi-layer perceptron (left). We show how to accelerate this architecture using the cooperative vectors extension to obtain up to $23 \times$ speedup compared to a compute based with fused multiply add (FMA) implementation (right).
  }
  \label{fig:teaser}
}

\maketitle
%-------------------------------------------------------------------------
\begin{abstract}
In this work, we present an extension to the neural texture compression method of Weinreich and colleagues~\cite{weinreich2024real}. Like them, we leverage existing block compression methods which permit to use hardware texture filtering to store a neural representation of physically-based rendering (PBR) texture sets (including albedo, normal maps, roughness, etc.). However, we show that low dynamic range block compression formats still make the solution viable. Thanks to this, we show that we can achieve higher compression ratio or higher quality at fixed compression ratio. We improve performance at runtime using a tile based rendering architecture that leverage hardware matrix multiplication engine. Thanks to all this, we render 4k textures sets (9 channels per asset) with anisotropic filtering at 1080p using only $28$MB of VRAM per texture set at $0.55$ms on an Intel B580.

%\begin{CCSXML}
%<ccs2012>
%   <concept>
%       <concept_id>10010147.10010371.10010395</concept_id>
%       <concept_desc>Computing methodologies~Image compression</concept_desc>
%       <concept_significance>500</concept_significance>
%       </concept>
%   <concept>
%       <concept_id>10010147.10010371.10010372</concept_id>
%       <concept_desc>Computing methodologies~Rendering</concept_desc>
%       <concept_significance>500</concept_significance>
%       </concept>
% </ccs2012>
%\end{CCSXML}
%
%\ccsdesc[500]{Computing methodologies~Image compression}
%\ccsdesc[500]{Computing methodologies~Rendering}
%
%
%\printccsdesc   

\end{abstract}  

%-------------------------------------------------------------------------
\section{Introduction}

Modern games rely increasingly on high-resolution textures (often $4096 \times 4096$ pixels) to store the appearance of assets. Since physically-based shading models require many parameters (albedo, normal map, roughness, metalness, \textit{etc})~\cite{hill2020physically} for their evaluation, many textures have to be stored for a single asset. This comes at the cost of storage capacity both on disk and on graphics memory (VRAM). In fact, some AAA games can go beyond $100$Gb~\cite{fowler202extending} on disk, with most of this storage dedicated to textures. To mitigate this, texture compression algorithm must be used. However, decompression will impact the performance either at runtime or during loading time. \change{Furthermore, decompression must be compatible with texture mipmapping for level-of-details.} Hence, hardware-accelerated compression methods such as block compression~\cite{khronos2025image} (BC1 to BC7) are the \textit{de facto} standard.
Still, this type of compression does not use the fact that different textures are used for the same asset and could be further compressed by leveraging the correlation between different parameters.

\paragraph*{Neural Material Compression.} Higher compression ratio through texture grouping is the key idea behind neural material compression. Instead of compressing textures individually, neural material compression methods~\cite{farhadzadeh2024neural,vaidyanathan2023random,weinreich2024real} stack together all textures belonging to the same asset (called a texture set) and compress this tensor into a lighter representation. A simple Multi-Layer Perceptron is used to decode this representation to an approximation of the original textures. Still, those methods trade lower footprint on disk and/or VRAM for compute since a neural network has to be evaluated.

\paragraph*{Our method.} Our work aims to improve the Block Compressed Features (BCF) method~\cite{weinreich2024real} both in terms of compression ratio, quality, as well as performance. More specifically, we
\begin{itemize}
    \item use BC1 in place of BC6 with a different resolution layout, allowing more compression or quality depending on the requirements (see Section~\ref{sec:method}).
    \item show that anisotropic filtering is possible with those models (see Section~\ref{sec:runtime-filtering}).
    \item reduce decoding time with hardware acceleration of matrix-vector multiplication made efficient thanks to a tile-based rendering architecture (leveraging the \textit{cooperative vector} extension) (see Section~\ref{sec:hardware-acceleration}).
    % \item \idea{Distillation of the model to support different horsepowers (going from $64$ hidden dim to $16$).}
    % \item \idea{Convert compressed representation directly to BC format.}
\end{itemize}

%-------------------------------------------------------------------------
\section{Previous Work}

Texture compression plays a crucial role in optimizing GPU memory usage and performance in modern graphics applications. Over the years, significant advancements have been made to improve compression efficiency, decoding speed, and visual quality while keeping memory requirements manageable (see the evolution of JPEG formats~\cite{wallace1992jpeg,christopoulos2000jpeg2000,alakuijala2019jpeg} for example). However, not all compression formats are usable in 3D graphics, as hardware accelerated decoding and filtering must be available.

\paragraph*{Texture Compression.}
The block Compression (BC) family of formats~\cite{delp1979image,franti1994compression}, (BC1 to BC7), is widely used in gaming and graphics due to its balance between compression ratio, visual quality, and decoding speed. These formats use fixed-size blocks (typically 4x4 pixels) and are optimized for GPU hardware, enabling efficient decompression during rendering.

ASTC~\cite{nystad2012adaptive} is a more flexible texture compression standard with adaptive range of block sizes, from $4 \times 4$ to $12 \times 12$ pixels, allowing for a finer control over the compression rate and visual quality. Despite its advantages, BC formats are often prefered over ASTC as they are widely supported.

\paragraph*{Neural Image Compression.}
Many work have leveraged the use of Machine Learning for image compression~\cite{balle2017endtoend}. However, most of these works are not usable in a real-time context, where fast decoding speed is mandatory.

Instead some works have focused on this specific challenge of compressing images with real-time usage. For example, Datta et al.~\cite{datta2023efficient} learn indirections to a buffer of values. Although it achieves good compression ratios, dependent fetches can be harmful to performance. Fujieda and Harada~\cite{fujieda2024neural} introduced Neural Texture Block Compression (NTBC), which uses neural networks to map uncompressed textures to compressed texture blocks. Their approach reduces disk storage costs by approximately $70\%$ while maintaining real-time performance. However, they decompress textures to BC format in VRAM.

\paragraph*{Neural PBR Material Compression.}
Instead of directly compressing images, some work leverages the correlation between the physically based materials channels. This can lead to higher compression ratios. Most notably, \cite{vaidyanathan2023random, farhadzadeh2024neural} use multiple grids and positional encoding to decode pixels. Those methods target high compression ratios, but require higher compute (2 hidden layers with 64 hidden channels) than alternatives.

Weinreich et al.~\cite{weinreich2024real} introduced a method that stores learned latent features as block-compressed high-dynamic range BC6 textures. Their approach leverages a small MLP decoder (1 hidden layer) that runs on the filtered latent features to infer the texture data. This allows for real-time decompression directly within shaders. We build on their method and show how to improve compression ratio and inference speed using LDR block compression and hardware accelerated inference on tiles.

%-------------------------------------------------------------------------
\section{Neural Block Texture Compression}
\label{sec:method}

In the following, we first describe our method's neural architecture (Section~\ref{sec:architecture}). Then, we detail how textures are compressed in our prototype (Section~\ref{sec:training}). We also discuss runtime evaluation and texture filtering (Section~\ref{sec:runtime-filtering}).

\subsection{Network Architecture}
\label{sec:architecture}

Figure~\ref{fig:teaser} displays the architecture of our model: a set of block compressed textures with different resolutions stores a latent representation. To evaluate a specific pixel at a given UV coordinate $p(\uv)$, we evaluate each latent texture $\cvec_k = T_k(\uv)$, concatenate all latent colors $\xvec = \left[\cvec_0~\dots~\cvec_K\right] $ and feed the resulting vector to a Multi-Layers Perceptron (MLP) to obtain a vector of features (albedo, normal, ...): $\fvec = f(\xvec)$.

\paragraph*{Latent BC1 compression.}
Each latent texture $T_k$ has its own resolution ($W_k \times H_k$) and is stored in BC1 format (see Khronos documentation~\cite{khronos2025image} for details). Hence, each pixel is defined as the interpolation of two endpoints $\mathbf{e}_0, \mathbf{e}_1$ with a blending factor $\alpha$:
$$ T(\uv) = \left( 1-\alpha(\uv) \right) \mathbf{e}_0(\uv)  + \alpha(\uv) \mathbf{e}_1(\uv).$$
With BC1 format, the endpoints are shared by groups of $4 \times 4$ texels while every texel has its own $\alpha$. In practice, $\alpha$ is quantized using 2 bits and $\mathbf{e}_0$, $\mathbf{e}_1$ are quantized with a $[5,6,5]$ bits pattern.

\paragraph*{Difference with BCF.} A key difference with the method of Weinreich \textit{et al.}~\cite{weinreich2024real} is that they use BC6 compression. Their main reason to use BC6 is that it allows to decode HDR values. However, we did not find it to be necessary in practice. Instead, we can store twice the amount of BC1 textures for the same footprint in BC6. In the following, we refer to BCF~\cite{weinreich2024real} as BCF6 and ours as BCF1\footnote{we could use other kind of image compression such as BC2, BC3, \textit{etc}. We did not test that in this article.}.

\paragraph*{Variants.} We tested two variants with four textures per material: \variantA~and \variantB. They differ by the resolution of the latent texture: we designed \variantA~to have almost the same compression ratio as BCF6 while \variantB~almost double this compression ratio.
\begin{center}
    \begin{tabular}{|c|cccc|}
    \hline
     \textbf{variant}    & \multicolumn{4}{c|}{\textbf{resolutions}}  \\
     \hline
     \variantA    & $W\times H$ & $W\times H$ & ${W \over 2}\times{H \over 2}$ & ${W \over 2}\times{H \over 2} $  \\
     \variantB    & $W\times H$ & ${W \over 2}\times{H\over 2}$& ${W\over 4}\times{H \over 4}$ & ${W\over 8}\times{H \over 8}$  \\
     \hline
    \end{tabular}
\end{center}

\begin{figure}[h]
    \centering
    \input{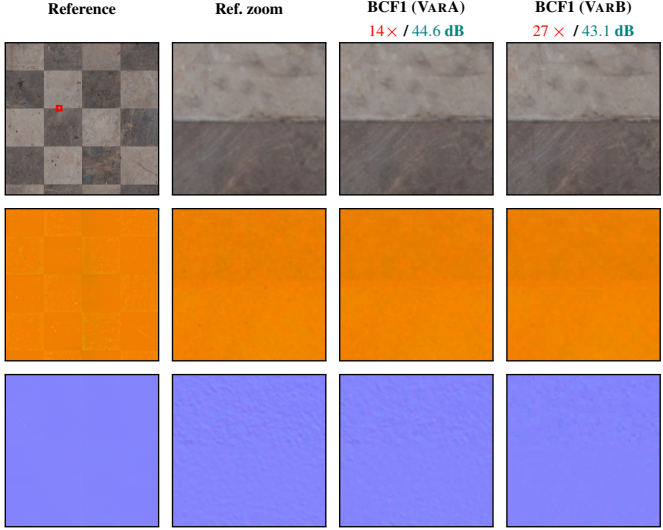}
    \caption{
        \textbf{Quality of the different variants.}
        Using \variantA~or \variantB~permit to adapt the compression level ({\color{red}in red}) with an impact on quality (average PSNR {\color{teal}in green}). As we double the compression ratio, the block artifacts of BC1 reduce the quality of the inference.
    }
    \vspace{-7pt}
    \label{fig:results_quality_variants}
\end{figure}

Furthermore, we apply sub-pixel shifts during the evaluation of the latent textures. We shift by half a texel the second and fourth textures. \change{This avoids the alignment of latent blocks to reduce artefacts}. We show in Figure~\ref{fig:results_quality_variants} the impact of both variants on image quality. See our supplemental material for more results.

\subsection{Training}
\label{sec:training}
To compress texture into our latent BC1 format, we replicated a BC1 texture unit in PyTorch~\cite{paszke2019pytorch}. Since BC1 texture store mip hierarchies as separate BC1 textures, we optimize them as well. During training, we sample a LOD level and regress the associated texture. We sample the LOD level similarly to Vaidyanathan~\cite{vaidyanathan2023random}.

We apply a L1 loss between the decoded features and reference features sampled using a trilinear filtering. Gradients are passed through the MLP and each element of the BC1 textures are updated using the Adam optimizer (with $lr=10^{-3}$ and $lr=10^{-2}$ respectively). For each BC1 texture, we store its endpoints and alpha as floating point tensor. To decompress the unsigned quantized values ($\bar{\alpha}$, $\bar{\mathbf{e}}_0$, and $\bar{\mathbf{e}}_1$), we apply a sigmoid activation function to the floating point values before performing quantization aware training:
\begin{align}
    \bar{\alpha}       &= \text{quant}\left(\text{sigmoid}\left( \alpha \right), [2] \right)\\
    \bar{\mathbf{e}}_0 &= \text{quant}\left(\text{sigmoid}\left( \alpha \right), [5,6,5] \right)\\
    \bar{\mathbf{e}}_1 &= \text{quant}\left(\text{sigmoid}\left( \alpha \right), [5,6,5] \right)
\end{align}
where $\text{quant}(x, b)$ evaluate the quantized form of $x$ using $b$ bits but let the gradients of its floating point value pass through.

We did not find evidence that optimizing first unquantized latent maps before quantizing during training was brining any gain in quality. Hence, we optimize our model with quantized representation from the start.

\subsection{Anisotropic Filtering}
\label{sec:runtime-filtering}
\begin{figure}[h]
    \centering
    \begin{tikzpicture}[font=\tiny]
        \node[inner sep=0pt] (iso_ours) { \includegraphics[width=0.245\linewidth]{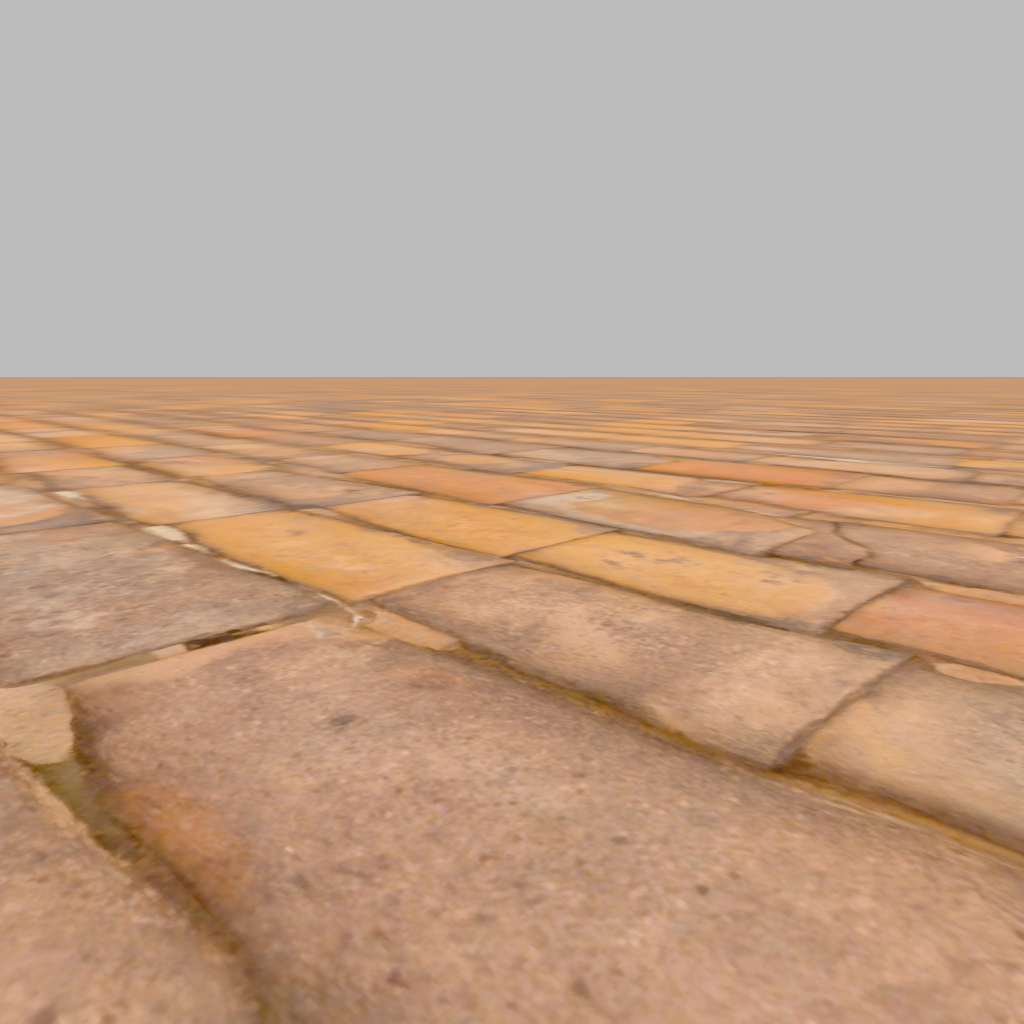}};
        \begin{scope}
            \clip (iso_ours.south west) rectangle (iso_ours.north);
            \node[inner sep=0pt, left=0pt of iso_ours, anchor=west] (iso_ref)  { \includegraphics[width=0.245\linewidth]{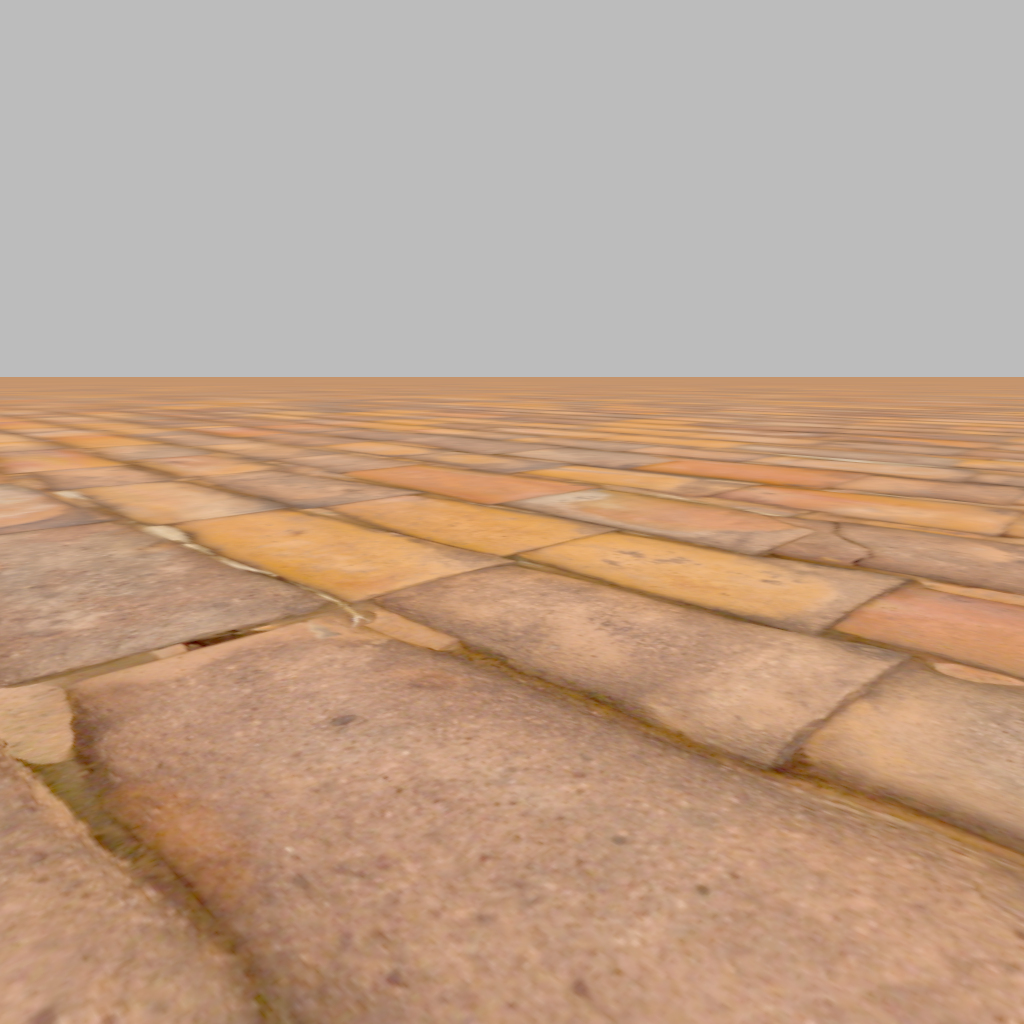}};
        \end{scope}
        \draw (iso_ours.south west) rectangle (iso_ours.north);
        \draw (iso_ours.south west) rectangle (iso_ours.north east);
        \node[below=0pt of iso_ours.north west, anchor=north west, minimum width=0.122\linewidth] { \textbf{BCF1} };
        \node[below=0pt of iso_ours.north east, anchor=north east, minimum width=0.122\linewidth] { \textbf{Ref.} };

        \node[inner sep=0pt, right=2pt of iso_ours] (iso_ours2) { \includegraphics[width=0.245\linewidth]{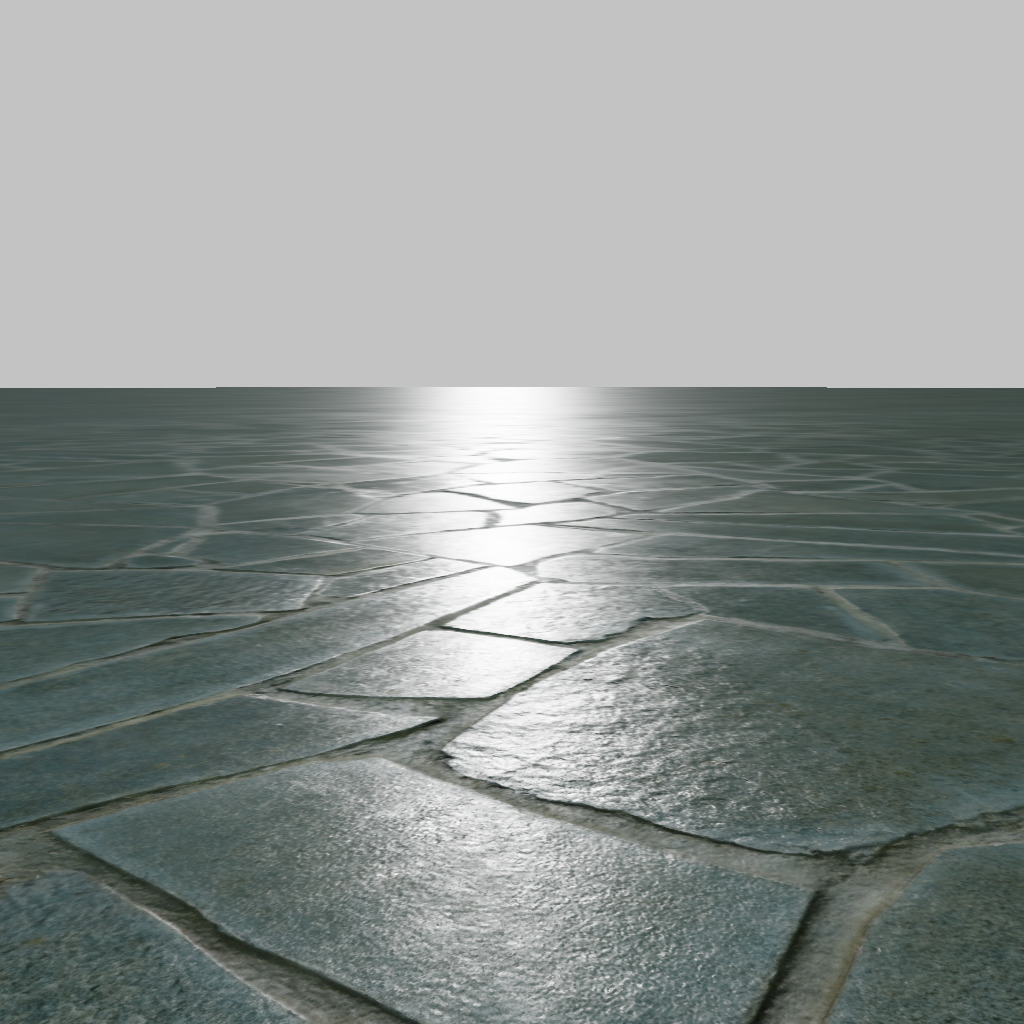}};
        \begin{scope}
            \clip (iso_ours2.south west) rectangle (iso_ours2.north);
            \node[inner sep=0pt, left=0pt of iso_ours2, anchor=west] (iso_ref2)  { \includegraphics[width=0.245\linewidth]{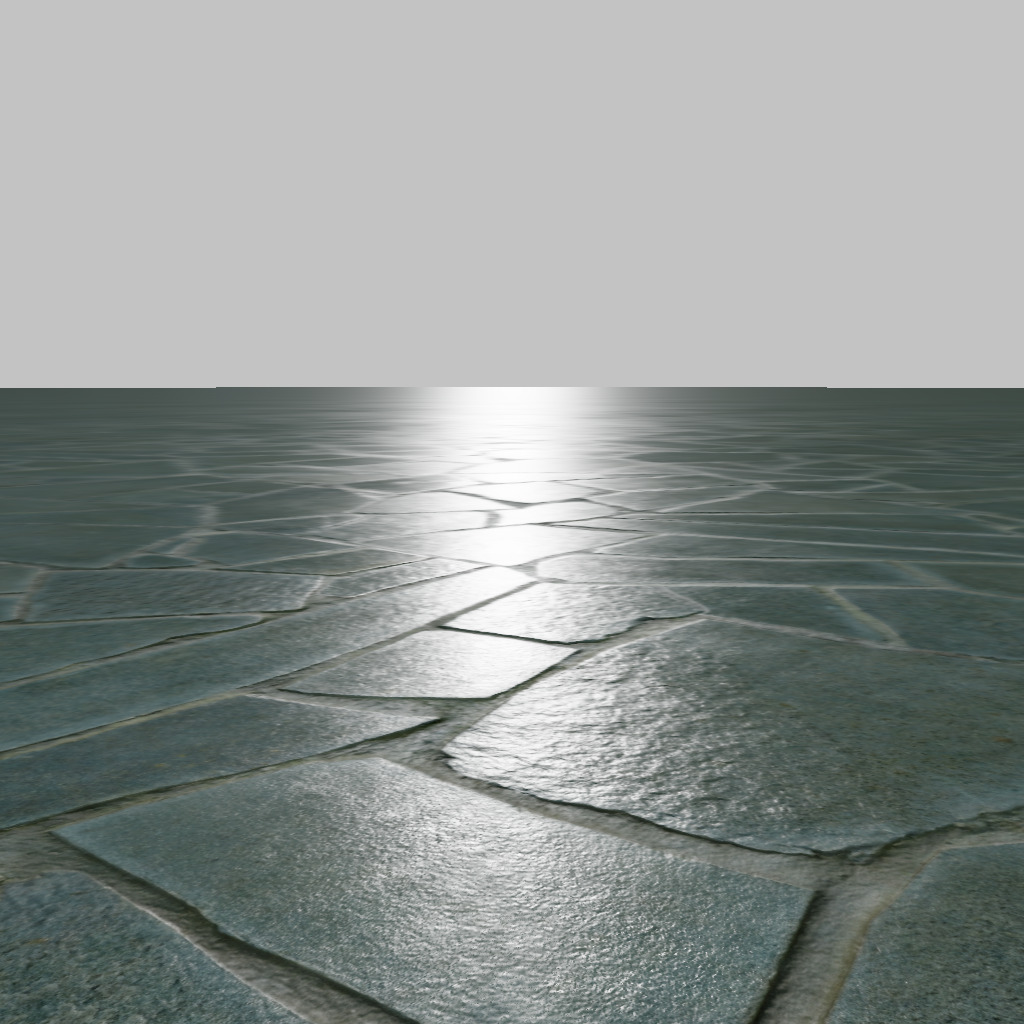}};
        \end{scope}
        \draw (iso_ours2.south west) rectangle (iso_ours2.north);
        \draw (iso_ours2.south west) rectangle (iso_ours2.north east);
        \node[below=0pt of iso_ours2.north west, anchor=north west, minimum width=0.122\linewidth] { \textbf{BCF1} };
        \node[below=0pt of iso_ours2.north east, anchor=north east, minimum width=0.122\linewidth] { \textbf{Ref.} };
        
        \draw[|-|] ($(iso_ours.north west)+(0,0.2)$)-- ($(iso_ours2.north east)+(0,0.2)$);
        \node[above=7pt of iso_ours.north east, anchor=center, fill=white] { \textbf{Isotropic filtering} };

        \node[inner sep=0pt, right=2pt of iso_ours2] (aniso_ours) { \includegraphics[width=0.245\linewidth]{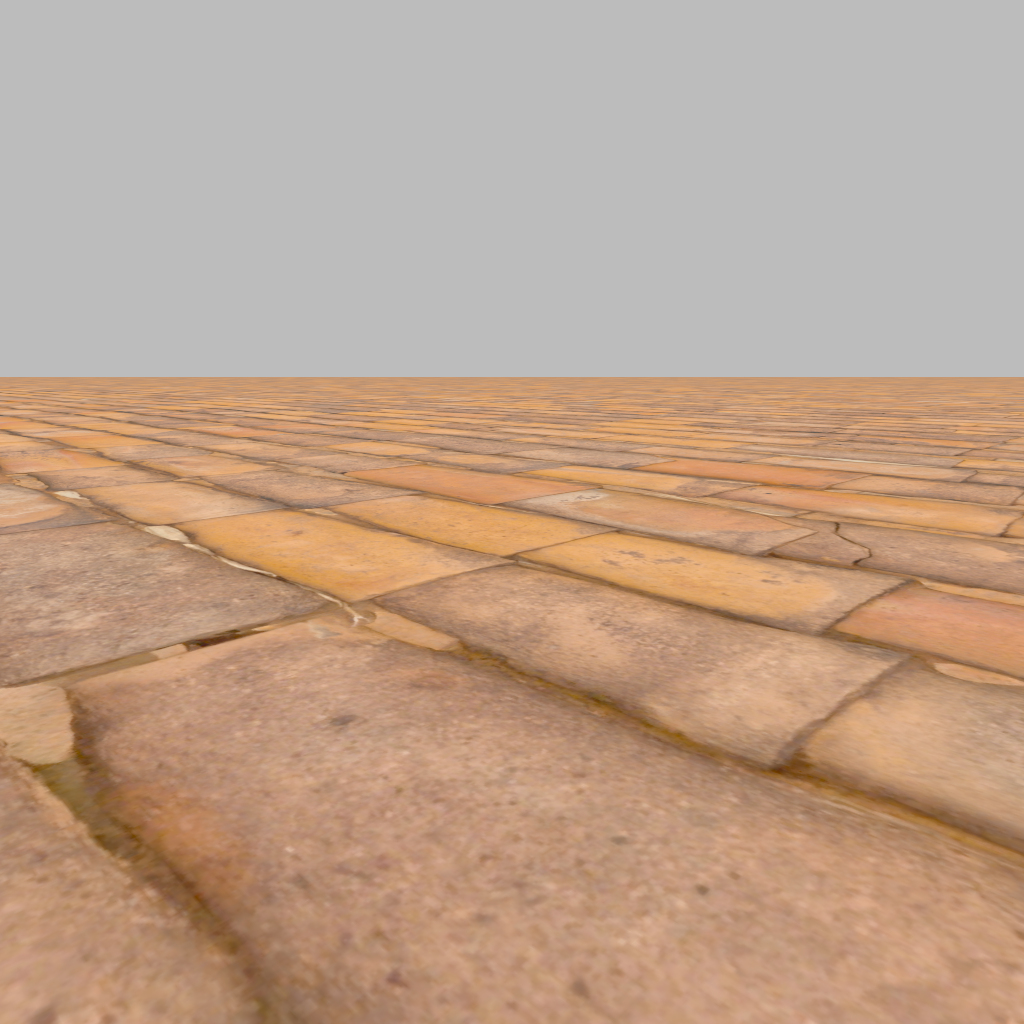}};
        \begin{scope}
            \clip (aniso_ours.south west) rectangle (aniso_ours.north);
            \node[inner sep=0pt, left=0pt of aniso_ours, anchor=west] (aniso_ref)  { \includegraphics[width=0.245\linewidth]{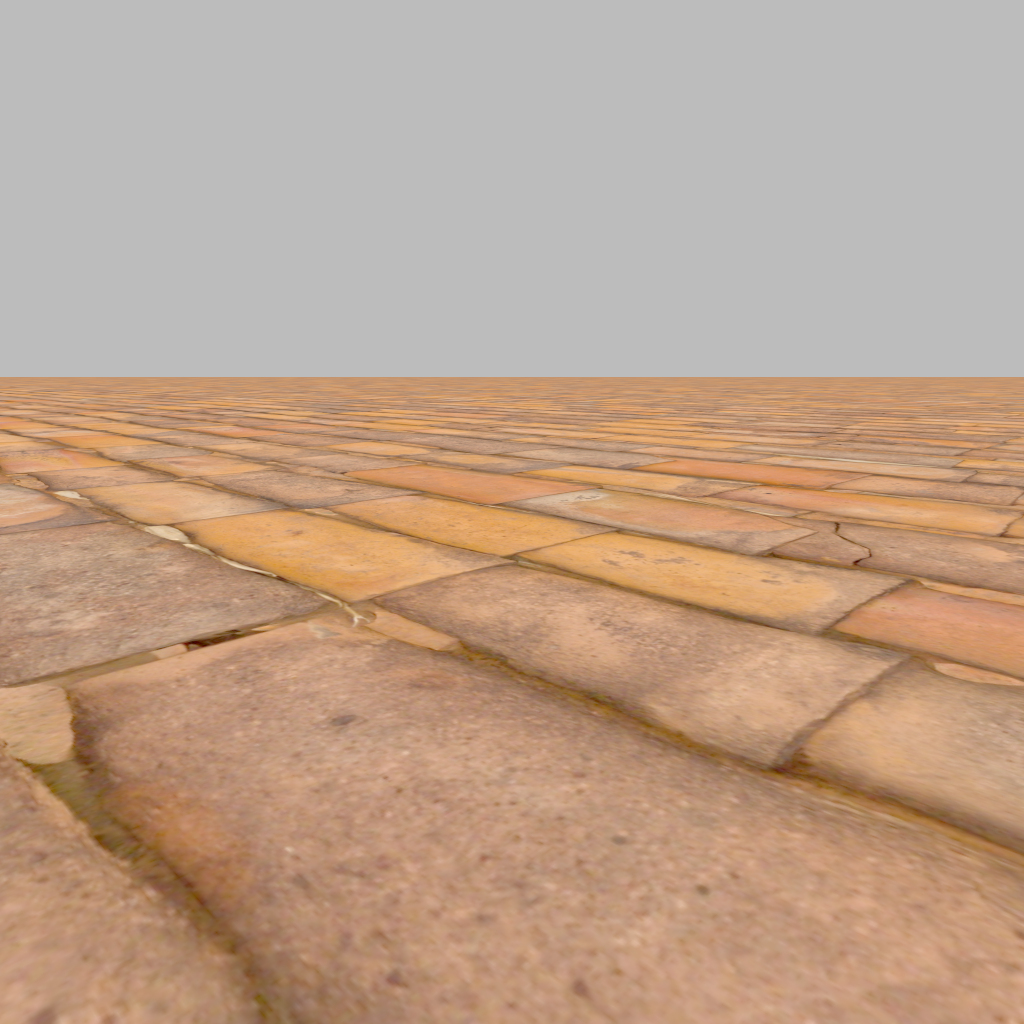}};
        \end{scope}
        \draw (aniso_ours.south west) rectangle (aniso_ours.north);
        \draw (aniso_ours.south west) rectangle (aniso_ours.north east);
        \node[below=0pt of aniso_ours.north west, anchor=north west, minimum width=0.122\linewidth] { \textbf{BCF1} };
        \node[below=0pt of aniso_ours.north east, anchor=north east, minimum width=0.122\linewidth] { \textbf{Ref.} };

        \node[inner sep=0pt, right=2pt of aniso_ours] (aniso_ours2) { \includegraphics[width=0.245\linewidth]{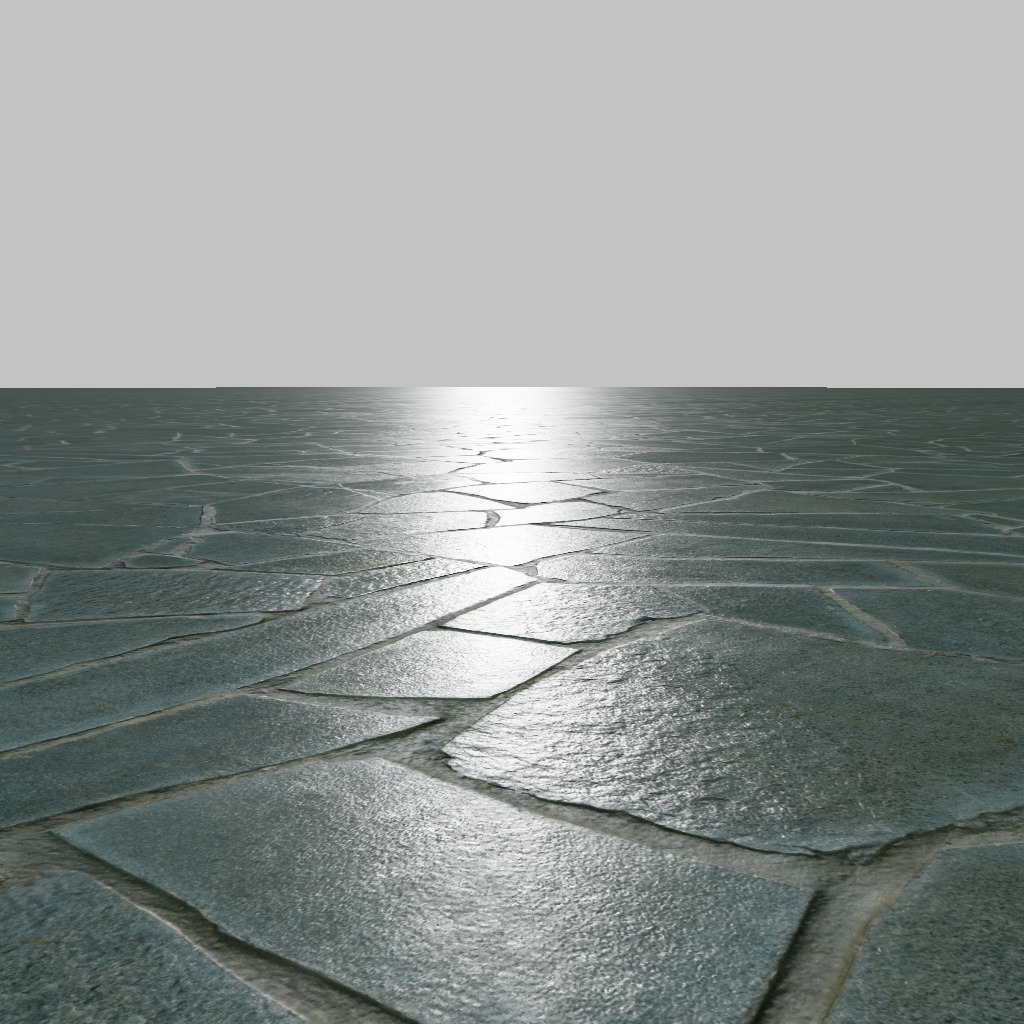}};
        \begin{scope}
            \clip (iso_ours2.south west) rectangle (aniso_ours2.north);
            \node[inner sep=0pt, left=0pt of aniso_ours2, anchor=west] (iso_ref2)  { \includegraphics[width=0.245\linewidth]{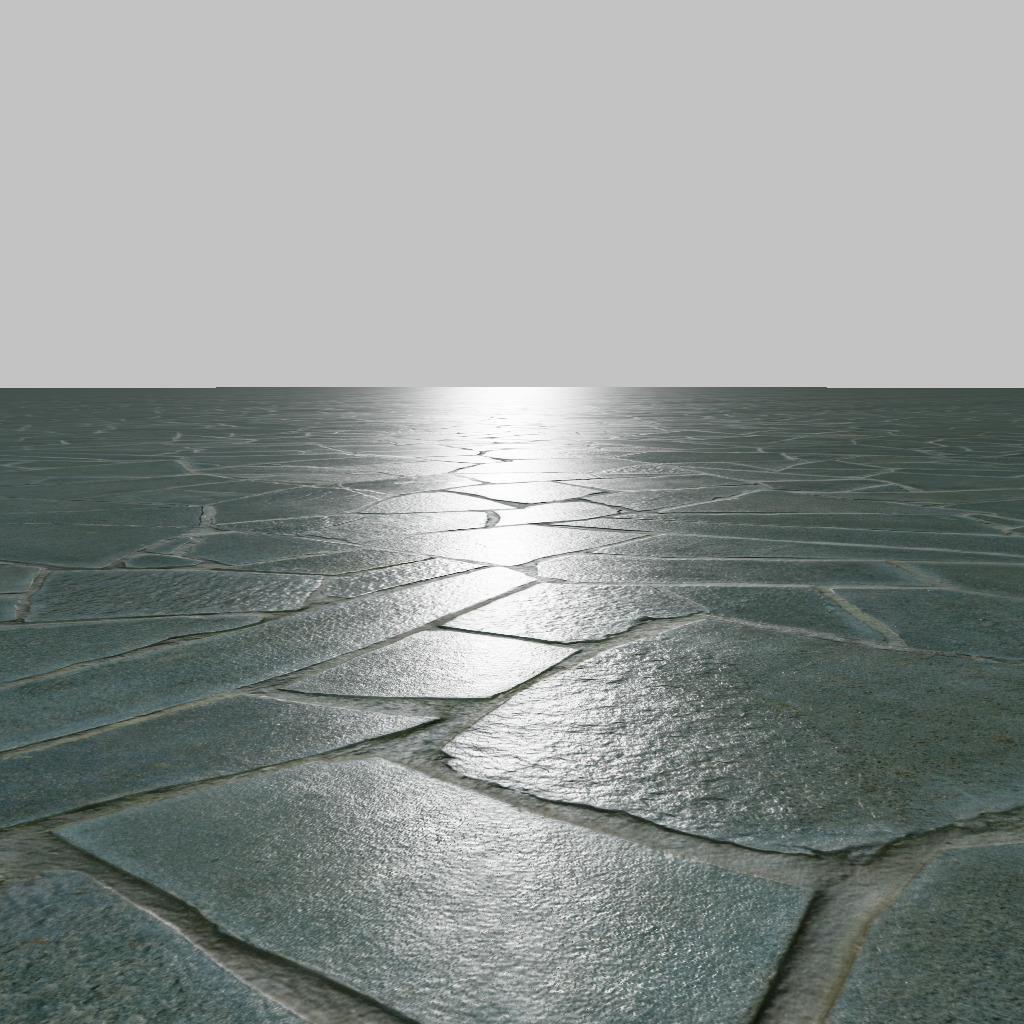}};
        \end{scope}
        \draw (iso_ours2.south west) rectangle (aniso_ours2.north);
        \draw (iso_ours2.south west) rectangle (aniso_ours2.north east);
        \node[below=0pt of aniso_ours2.north west, anchor=north west, minimum width=0.122\linewidth] { \textbf{BCF1} };
        \node[below=0pt of aniso_ours2.north east, anchor=north east, minimum width=0.122\linewidth] { \textbf{Ref.} };

        \draw[|-|] ($(aniso_ours.north west)+(0,0.2)$)-- ($(aniso_ours2.north east)+(0,0.2)$);
        \node[above=7pt of aniso_ours.north east, anchor=center, fill=white] { \textbf{Anisotropic filtering} };
    \end{tikzpicture}
    % \vspace{-20pt}
    \caption{
        \change{\textbf{Isotropic and anisotropic fitlering.} We validate that training with isotropic filtering permits to also extrapolate anisotropic filtering in our \textsc{VarA}.}
    }
    \label{fig:results_anisoptric_filtering}
    \vspace{-20pt}
\end{figure}

We found that using such optimization provides good results with anisotropic filtering when the latent texture are fetched with non-isotropic kernels. While our models are never trained on anisotropicaly filtered data, they still provide visualy correct anisotropic filtering without any change. We hypothetize that since the network has learned to infer the behaviour of bilinearly filtered latents, as a side effect it learned that blending latent codes results in blended outputs. Since anistropic filtering is a blend of many taps, this behavior is not surprizing. In Figure~\ref{fig:results_anisoptric_filtering}, we compare renderings of our decompressed textures with respect to the reference using isotropic or anisotropic filtering for both.

\section{Runtime \& Hardware Acceleration}
\label{sec:hardware-acceleration}
Given that the neural texture set is stored in the BC1 format and leverages the texture unit, the main bottleneck in the runtime evaluation lies in how the MLP is evaluated and where in the graphics pipeline it is evaluated.

\paragraph*{Rendering architecture.}
In our prototype, we implemented a rendering pipeline built around a \textit{Visibility buffer}~\cite{burns2013visibility}. This does mimick the integration of our method in a modern game engine. The \textit{Visibility pass} is followed by a \textit{Classification pass} that sort pixels requiring neural inference. Then, a \textit{G-Buffer pass} consumes the BCF1 texture set and produces all the data required for shading (albedo, normal, roughness, \textit{etc}). Finally, a \textit{Lighting pass} consumes the GBuffer and shades each pixel. We implemented a simple shading model that blends a diffuse lobe and a GGX lobe using prefiltered Image Based Lighting and a directional light~\cite{lagarde2015moving}.

\paragraph*{Neural inference in G-Buffer pass.}
We use the texture unit for an efficient fetch of block compressed latent features and unless noted, we use hardware accelerated anisotropic filtering. To improve performance in the \textit{G-Buffer pass}, we  take advantage of the hardware matrix multiplication engines to accelerate the inference of the differents layers of the Multi-Layer Perceptron (MLP). Our implementation use the \texttt{cooperative vectors} extension~\cite{bolz2024cooperative,cassie2024cooperative}. This provides better performance than using the \texttt{cooperative matrix} API or using software FMA for matrix multiplication. In our use of the cooperative vectors, we only use the column major layout and let the optimized opaque layout for future tests. But for \textit{cooperative vectors} to be efficient, threads in the same workgroup must use the same base matrix. Hence, we need to group together pixels using the same MLP.

\begin{figure}[t!]
    \begin{tikzpicture}[font=\tiny]
        % \draw (0,0) rectangle (\linewidth,2.4);
        \node[inner sep=0pt, anchor=south west] (classif) at (0,0) { \includegraphics[width=0.4\linewidth,trim={5.0cm 0 5.0cm 0},clip]{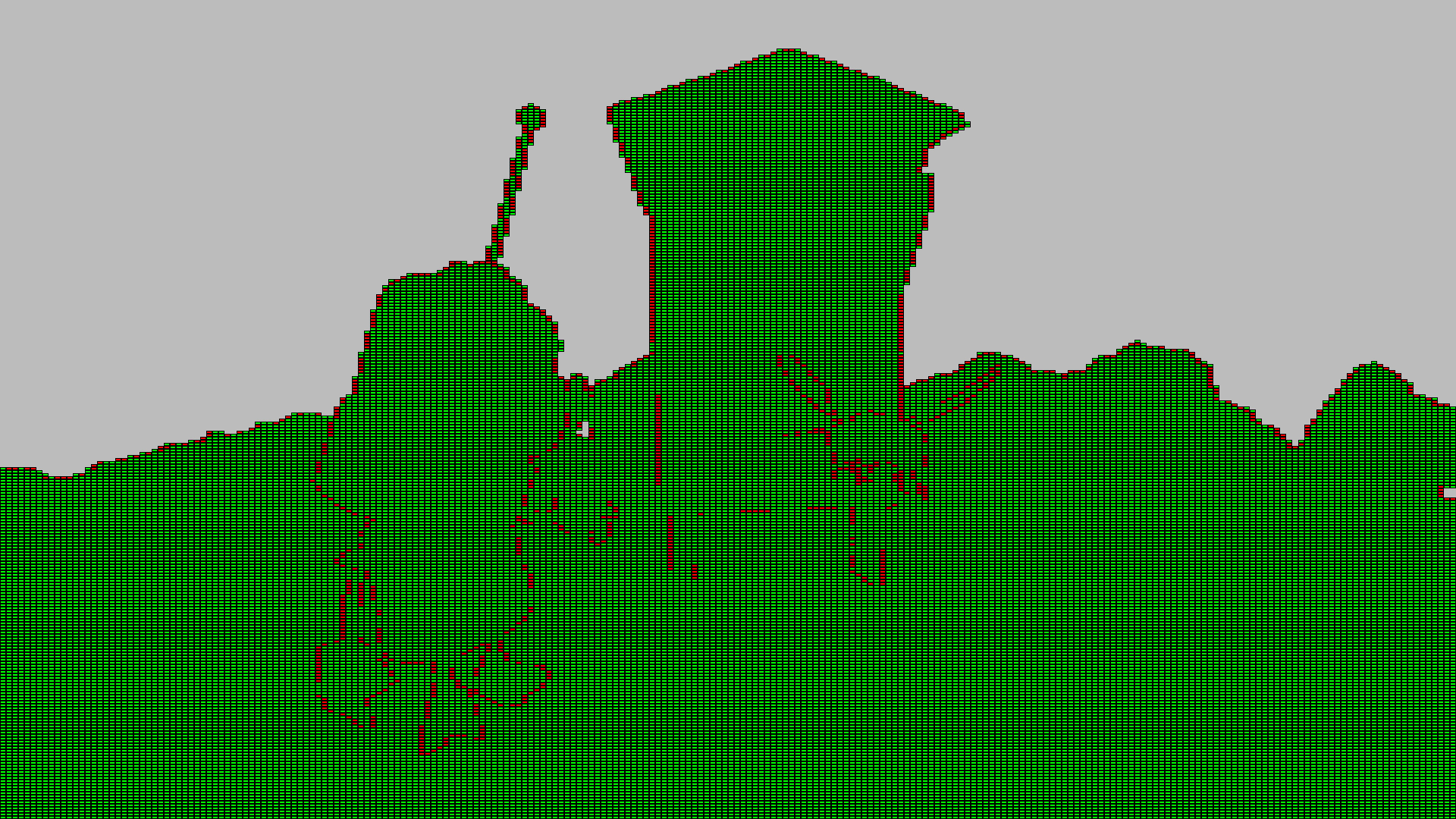}};
        \draw (classif.south west) rectangle (classif.north east);

        \node (L) at (1.0\linewidth,1.1cm) {Lighting};
        % \draw[fill, color=red] (classif)++(0.5, 0.0) rectangle +(0.5cm, 0.5cm);
        % \draw (0,0) rectangle +(1,1);

        \draw[->] (classif)++(0.9, +0.9) -- +(0.2\linewidth - 0.4cm, 0.0) node[inner sep=0pt] (A) {};
        \draw[fill=gray, thin] (classif)++(0.9, +0.9) ++(-0.025,-0.05) rectangle ++(0.05,0.1);
        \draw[fill=gray, thin] (A) ++(0.1,-0.2) rectangle ++(0.2,0.4);
        % \draw[ultra thin] (A) ++(0.1,-0.2) grid [step=0.05cm] ++(0.2,0.4);
        \begin{scope}[shift={($(A)+(0.1,-0.2)$)}]
            \draw[ultra thin] (0,0) grid [step=0.05cm] ++(0.2,0.4);
        \end{scope}
        \draw[->] (A)++(0.4, 0.0) -- ++(0.10\linewidth, 0.0) node[anchor=west, fill=white] { {\tiny no MLP } }-- ++(0.28\linewidth, 0.0) |- (L);

        \draw[->] (classif)++(0.5,  0.0) -- ++(0.2\linewidth, 0.0) node[inner sep=0pt] (B) {};
        \draw[fill=red, thin] (classif)++(0.5, +0.0) ++(-0.025,-0.05) rectangle ++(0.05,0.1);
        \draw[fill=red] (B) ++(0.1,-0.2) rectangle ++(0.2,0.4);
        % \draw[ultra thin] (B) ++(0.1,-0.2) grid [step=0.05cm] ++(0.2,0.4);
        \begin{scope}[shift={($(B)+(0.1,-0.2)$)}]
            \draw[ultra thin] (0,0) grid [step=0.05cm] ++(0.2,0.4);
        \end{scope}

        \draw[fill=orange] (B) ++(0.8,+0.3) node (B1a) {} rectangle +(0.1,0.2) ++(0.2,0.0) rectangle +(0.1,0.2) ++(0.2,0.0) rectangle +(0.1,0.2) ++(0.15,0.1) node[anchor=west, inner sep=0pt] {$\cdots$} ++(0.4,0.0) ++(0.0,-0.1) rectangle ++(0.1,0.2) node (B1) {};
        \draw[fill=olive] (B) ++(0.8,+0.0) node (B2a) {} rectangle +(0.1,0.2) ++(0.2,0.0) rectangle +(0.1,0.2) ++(0.2,0.0) rectangle +(0.1,0.2) ++(0.15,0.1) node[anchor=west, inner sep=0pt] {$\cdots$} ++(0.4,0.0) ++(0.0,-0.1) rectangle ++(0.1,0.2) node (B2) {};
        \draw[fill=teal] (B) ++(0.8,-0.4) node (B3a) {} rectangle +(0.1,0.2) ++(0.2,0.0) rectangle +(0.1,0.2) ++(0.2,0.0) rectangle +(0.1,0.2) ++(0.15,0.1) node[anchor=west, inner sep=0pt] {$\cdots$} ++(0.4,0.0) ++(0.0,-0.1) rectangle ++(0.1,0.2) node (B3) {};

        \draw[->] (B1)++(0.1, -0.1) -- ++(0.05\linewidth, 0.0) node[anchor=west] (B1k){ {\tiny MLP $1$ } };
        \draw[->] (B2)++(0.1, -0.1) -- ++(0.05\linewidth, 0.0) node[anchor=west] (B2k){ {\tiny MLP $2$ } };
        \draw[->] (B3)++(0.1, -0.1) -- ++(0.05\linewidth, 0.0) node[anchor=west] (B3k){ {\tiny MLP $N$ } };

        \draw[->] (B1k)++(0.4, 0.0) -- ++(0.398,0.0) |- (L);
        \draw[->] (B2k)++(0.4, 0.0) -- ++(0.398,0.0) |- (L);
        \draw[->] (B3k)++(0.4, 0.0) -- ++(0.375,0.0) |- (L);

        \draw (B)++(0.20,0.11) |- ($(B1a.north west) + (0.15,0.0)$);
        \draw (B)++(0.15,0.0) -- ++(0.4,0.0) |-($(B2a.north west) + (0.18,-0.05)$);
        \draw (B)++(0.25,-0.11) |- ($(B3a.north west) + (0.15,-0.08)$);

        \draw[->] (classif)++(0.0, -0.9) -- ++(0.2\linewidth + 0.5cm, 0.0) node[inner sep=0pt] (C) {};
        \draw[fill=green, thin] (classif)++(0.0, -0.9) ++(-0.025,-0.05) rectangle ++(0.05,0.1);
        \draw[fill=green] (C) ++(0.1,-0.2) rectangle ++(0.2,0.4);
        % \draw[ultra thin] (C) ++(0.1,-0.2) grid [step=0.05cm] ++(0.2,0.4);
        \begin{scope}[shift={($(C)+(0.1,-0.2)$)}]
            \draw[ultra thin] (0,0) grid [step=0.05cm] ++(0.2,0.4);
        \end{scope}
        \draw[->] (C)++(0.4, 0.0) -- ++(0.23\linewidth, 0.0) node[anchor=west] (Ck) { {\tiny MLP $k$ } };
        \draw[->] (Ck)++(0.4, 0.0) -- ++(0.435,0.0) |- (L);

    \end{tikzpicture}
    \caption{\textbf{Tile-based classification.} We tile the output of the \textit{Visibility pass} in $8 \times 4$ pixels tiles. Each tile is classified as either \textit{non neural} (gray), \textit{neural with a single MLP} (green), and \textit{neural with mixed MLPs} (red). To avoid divergence, we repack all \textit{mixed} tiles to have $8\times 4$ groups of pixels with the same MLP (orange, marron, blue). Once this sorting is done, we can perform our \textit{Lighting pass}. %Repacking group inference to cooperative vectors workgroup size (16 threads) improves performance.
    \label{fig:tile_based_classification}
    \vspace{-27pt}
    }
\end{figure}

\paragraph*{Tile-based classification.}
For improved performance, we leverage a tile-based architecture where both the \textit{G-Buffer} and \textit{Lighting} passes operate on $8 \times 4$ pixels tiles to match workgroups size. During a \textit{Classification pass}, we classify each $8 \times 4$ tile based on its content so as to perform neural decompression only on tiles that have at least one pixel using a neural texture set (see Figure~\ref{fig:tile_based_classification}). We detail this in Algorithm~\ref{alg:tile_based_classification}. We first classify all tiles into three types (\texttt{classificationA} in Algorithm~\ref{alg:tile_based_classification}): \textit{no neural} for tiles that do not require neural decompression; \textit{simple neural} for tiles with the same network for all its pixels; and \textit{mixed neural} for tiles with at least two different networks. We then perform an additional compute pass on \textit{mixed neural} tiles and sort their pixels by their MLP index (\texttt{classificationB} in Algorithm~\ref{alg:tile_based_classification}). This groups pixels with the same MLP in $8 \times 4$ tiles on which we execute the \textit{Lighting pass} before splatting the result in the associated pixel.

%-------------------------------------------------------------------------
% \vspace{-5pt}
\section{Results}

We tested both the quality and performance of our method. We refer to our supplemental material and video for more results. In all cases, we compressed PBR textures using a PyTorch implementation of the compressor.

\paragraph*{Quality and compression.}
We tested the quality and compression factor of our method using a dataset of texture sets from Polyhaven~\cite{polyhaven}. For every texture set, we selected the diffuse, normal, roughness, metalness and ambient occlusion channels, resulting in texture sets with 9 channels. We used $4096 \times 4096$ pixels textures in every cases.

We analyzed the quality (using the PSNR in decibels) for our different variants of BCF1. In Table~\ref{tbl:results_mlp_shape}, we report the different configurations we tested as well as the min, max, and average PSNR (in decibel). We found that using wider MLP is beneficial for quality. In the following, unless noted, we report results for $D=64$.

\begin{table}[h!]
    \center
    \begin{tabular}{|c|c|c|c|}
        \hline
        \textbf{variant} & \textbf{\numlayers~layers} & \textbf{hidden dim} & \textbf{PSNR}\\
        \hline\hline
        \multirow{3}{2em}{ \textbf{varA} }
        & $\numlayers=1$ & $D=16$ & $22.22 ~/~ 40.37 ~/~ 49.67$ \\
        & $\numlayers=1$ & $D=32$ & $22.70 ~/~ 40.85 ~/~ \mathbf{50.16}$ \\
        & $\numlayers=1$ & $D=64$ & $\mathbf{23.34} ~/~ \mathbf{40.91} ~/~ 50.02$ \\
        \hline
        \multirow{3}{2em}{ \textbf{varB} }
        & $\numlayers=1$ & $D=16$ & $22.11 ~/~ 38.28 ~/~ 48.74$ \\
        & $\numlayers=1$ & $D=32$ & $22.34 ~/~ 38.75 ~/~ 49.11$ \\
        & $\numlayers=1$ & $D=64$ & $\mathbf{22.91} ~/~ \mathbf{39.02} ~/~ \mathbf{49.18}$ \\
        \hline
    \end{tabular}
    \caption{
        \textbf{Dimension of the MLP.}
        We compressed textures from the Polyhaven texture dataset and report the quality in PSNR for different configurations of our variants.
        \label{tbl:results_mlp_shape}
    }
    % \vspace{-5pt}
\end{table}

\begin{figure}[t!]
    \centering
    \input{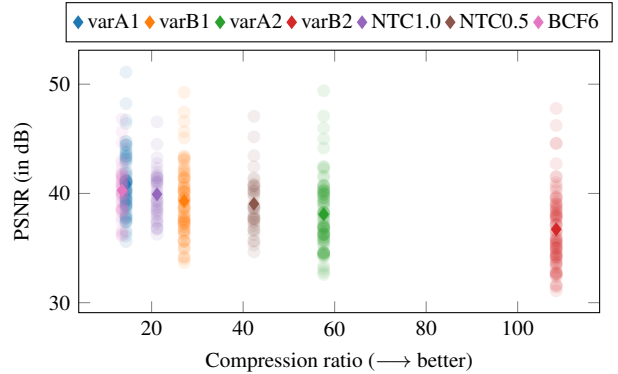}
    \vspace{-8pt}
    \caption{
        \textbf{Quality and compression ratio.}
        We compare the quality of our variants, varA and varB with latents at full resolution (varA1, varB1) and half resolution (varA2, varB2) respectively. We also plot the quality obtain using Nvidia's NTC (both 0.5 and 1.0 variants) and BCF using BC6H compression.
    }
    \vspace{-20pt}
    \label{fig:results_quality_vs_compression}
\end{figure}

We report as well the quality w/r/t the compression ratio of our variants and compare them to Ubisoft's BCF6 and Nvidia's NTC ($1$ and $\nicefrac{1}{2}$ bits per channel per pixel variants). Both BCF6 and NTC are our implementation, following the respective articles. For BCF6, we used the same mode for all blocks (mode 10 with RGB encoded with 6 bits per channel). We report the average PSNR and compression ratio in Figure~\ref{fig:results_quality_vs_compression}. Compared to BCF6, our BCF1 varA method produce similar quality with a small gain in compression. 

% \begin{itemize}
%     \item \idea{Peut-on avoir un seul MLP pour toutes les textures. \textbf{Laurent}}
%     \item \idea{Peut-on trouver une metrique qui permette de prédire la qualité de la compression en amont. \textbf{Laurent}}
% \end{itemize}

% \begin{figure}[t!]
%     \centering
%     \begin{tikzpicture}[font=\footnotesize]
%         \node at (0.0,0) {\includegraphics[width=0.25\linewidth]{./figs/magnification/aniso_uncompressed.jpg}};
%         \node at (2.5,0) {\includegraphics[width=0.25\linewidth]{./figs/magnification/aniso_neural.jpg}};
%     \end{tikzpicture}
%     \vspace{-8pt}
%     \caption{
%         \textbf{Quality and compression ratio.}
%         We compare the quality of our variants, varA and varB with latents at full resolution (varA1, varB1) and half resolution (varA2, varB2) respectively. We also plot the quality obtain using Nvidia's NTC (both 0.5 and 1.0 variants) and BCF using BC6H compression.
%     }
%     \vspace{-10pt}
%     \label{fig:results_quality_vs_compression}
% \end{figure}

\definecolor{nvidia}{HTML}{76B900}
\definecolor{intel}{HTML}{0068B5}

\begin{table*}[t!]
    \center
    \begin{tabular}{|c|c|c|c|c|c|}
        \hline
        \multicolumn{2}{|c|}{} & \textbf{no neural} &  \textbf{software (FMA)} & \textbf{coop vectors} & \textbf{PSNR}\\
        \hline\hline
        \multirow{3}{2em}{ \textbf{varA} }
        & $D=16$ & \multirow{4}{7em}{ \textcolor{nvidia}{$0.035$} / \textcolor{intel}{$0.11$} \textit{ms} } & \textcolor{nvidia}{$0.085$} / \textcolor{intel}{$0.183 $} \textit{ms} &  \textcolor{nvidia}{$0.085$} / \textcolor{intel}{$0.077 $}     \textit{ms}  ~~ ($\times ~1 - ~2$) & $34.70 ~/~ 37.49 ~/~ 38.52$ \\
        & $D=32$ &                                                                                          & \textcolor{nvidia}{$0.785$} / \textcolor{intel}{$0.388 $} \textit{ms} &  \textcolor{nvidia}{~~~$0.07$}  / \textcolor{intel}{$0.078 $}  \textit{ms}  ~~ ($\times ~5 - 11$) & $38.62 ~/~ 41.20 ~/~ 43.82$ \\
        & $D=64$ &                                                                                          & \textcolor{nvidia}{$2.8$}   / \textcolor{intel}{$2.3 $}   \textit{ms} &  \textcolor{nvidia}{~~~~$0.06$}  / \textcolor{intel}{$0.113 $} \textit{ms}  ~~ ($\times 20 - 46$) & $\mathbf{38.96} ~/~ \mathbf{41.76} ~/~ \mathbf{44.43}$ \\
        \hline
        \multirow{3}{2em}{ \textbf{varB} }
        & $D=16$ & \multirow{4}{7em}{ \textcolor{nvidia}{$0.035$} / \textcolor{intel}{$0.11$} \textit{ms} }  & \textcolor{nvidia}{$0.08$} / \textcolor{intel}{$0.183$} \textit{ms} &  \textcolor{nvidia}{~$0.085$} / \textcolor{intel}{$0.075 $}   \textit{ms}  ~~ ($\times ~1 - ~2$) & $34.97 ~/~ 35.83 ~/~ 36.83$ \\
        & $D=32$ &                                                                                           & \textcolor{nvidia}{$0.6$}  / \textcolor{intel}{$0.387$} \textit{ms} &  \textcolor{nvidia}{~~~~~~$0.06$}  / \textcolor{intel}{$0.079 $} \textit{ms} ~~~($\times 5 - 10$) ~~~~& $\mathbf{36.87} ~/~ 39.55 ~/~ \mathbf{43.22}$ \\
        & $D=64$ &                                                                                           & \textcolor{nvidia}{$2.7$}  / \textcolor{intel}{$2.3 $}  \textit{ms} &  \textcolor{nvidia}{~~$0.055$} / \textcolor{intel}{$0.11 $}    \textit{ms} ~~~~ ($\times 20-49$) & $36.86 ~/~ \mathbf{39.86} ~/~ 43.20$ \\
        \hline
    \end{tabular}
    \caption{
        \textbf{Performance and quality of our variants.}
        We measured timings of our prototype with (coop vectors) and without (software FMA) using hardware matrix multiplication on the Polyhaven assets database. Depending on the dimension of the MLP, hardware acceleration can provide a benefit. We also report the \textit{min} /  \textit{average} /  \textit{max} PSNR for each variant. We measured timings on an \textcolor{intel}{Intel B580 (in blue)} as well as an \textcolor{nvidia}{Nvidia RTX 4090 (in green)} GPUs.
        \label{tbl:results_tables}
        \vspace{-20pt}
    }
\end{table*}

\paragraph*{Performance.}
We implemented our method using the DirectX~12 API following Section~\ref{sec:hardware-acceleration}. Figure~\ref{fig:teaser} shows our prototype with a scene composed of $4$ assets (a robot, a bee, a bee hive, and rocks) each with its own neural material at $4096 \times 4096$ resolution. We tested our implementation on an Intel B580 and an Nvidia RTX 4090 GPUs. We set the screen resolution at $1920 \times 1080$ pixels for every result. Please refere to our supplemental video for more results.

We report performance measurements in Table~\ref{tbl:results_tables}. For every result, we skip the \textit{Visiblity} and \textit{Lighting pass} from the timings to focus on the impact of materials evaluation. We compare the impact of computing matrix multiplication in software using FMA versus leveraging the hardware acceleration using the cooperative vector API. When using wider MLPs ($D >= 32$), using cooperative vectors permits up to a $\times 40$ speedup.
% \begin{itemize}
%     \item \idea{Tester l'impact d'avoir 1 MLP different par pixel (cas extreme) ou bien plein de triangles avec chacun 1 MLP VS avoir 1 seul MLP. \textbf{Anis}}
% \end{itemize}

%-------------------------------------------------------------------------
\vspace{-5pt}
\section{Limitations}

\paragraph*{Scaling issue}
Our classification permits to handle different number of MLPs but is not yet optimal. We currently use global atomics to affect pixels in mixed tile to uniform tiles. This works fine when a small percentage of tiles are mixed, but shared memory's atomics would be more suited if all the screen is filled with mixed tiles.

\change{
\paragraph*{Sharing of MLP.} One possibility is to share a unique MLP for different materials, hence reducing the number of mixed tiles. This would improve performances at the expense of quality. We leave this feature for future works.
}

\section{Conclusion}
We have shown that using hardware acceleration provides a substancial benefit for the evaluation of neural materials. Furthermore, using \textit{low dynamic range} latent representations (using BC1) permits to obtain similar quality than previous work using \textit{high dynamic range} storage at the same compression rate and attain higher compression rate with a reasonnable quality drop.

We have shown that our compression approach when used in conjunction with coop vectors presents a viable alternative to traditional compression. We believe that standardizing a neural format will make neural texture compression even more appealing and thus constitutes a valuable future work.

%-------------------------------------------------------------------------

%\bibliographystyle{eg-alpha}
\bibliographystyle{eg-alpha-doi}

\bibliography{egbibsample}

\newcommand{\etalchar}[1]{$^{#1}$}
\begin{thebibliography}{\uppercase{WDOHN24}}

\bibitem[AVAB{\etalchar{*}}19]{alakuijala2019jpeg}
\textsc{Alakuijala J., Van~Asseldonk R., Boukortt S., Bruse M., Comșa I.-M.,
  Firsching M., Fischbacher T., Kliuchnikov E., Gomez S., Obryk R., et~al.}:
\newblock Jpeg xl next-generation image compression architecture and coding
  tools.
\newblock In \emph{Applications of digital image processing XLII} (2019),
  vol.~11137, SPIE, pp.~112--124.

\bibitem[BH13]{burns2013visibility}
\textsc{Burns C.~A., Hunt W.~A.}:
\newblock The visibility buffer: A cache-friendly approach to deferred shading.
\newblock \emph{Journal of Computer Graphics Techniques (JCGT) 2}, 2 (August
  2013), 55--69.
\newblock URL: \url{http://jcgt.org/published/0002/02/04/}.

\bibitem[BLS17]{balle2017endtoend}
\textsc{Ball{\'e} J., Laparra V., Simoncelli E.~P.}:
\newblock End-to-end optimized image compression.
\newblock In \emph{International Conference on Learning Representations}
  (2017).
\newblock URL: \url{https://openreview.net/forum?id=rJxdQ3jeg}.

\bibitem[cas25]{cassie2024cooperative}
Enabling neural rendering in directx: Cooperative vector support coming soon,
  2025.
\newblock URL:
  \url{https://devblogs.microsoft.com/directx/enabling-neural-rendering-in-directx-cooperative-vector-support-coming-soon/}.

\bibitem[CES00]{christopoulos2000jpeg2000}
\textsc{Christopoulos C.~A., Ebrahimi T., Skodras A.~N.}:
\newblock Jpeg2000: the new still picture compression standard.
\newblock In \emph{Proceedings of the 2000 ACM workshops on Multimedia} (2000),
  pp.~45--49.

\bibitem[DM79]{delp1979image}
\textsc{Delp E., Mitchell O.}:
\newblock Image compression using block truncation coding.
\newblock \emph{IEEE transactions on Communications 27}, 9 (1979), 1335--1342.

\bibitem[DMD{\etalchar{*}}23]{datta2023efficient}
\textsc{Datta S., Marshall C., Dong Z., Li Z., Nowrouzezahrai D.}:
\newblock Efficient graphics representation with differentiable indirection.
\newblock In \emph{SIGGRAPH Asia 2023 Conference Papers} (2023), pp.~1--10.

\bibitem[FH24]{fujieda2024neural}
\textsc{Fujieda S., Harada T.}:
\newblock Neural texture block compression.
\newblock \emph{arXiv preprint arXiv:2407.09543} (2024).

\bibitem[FHL{\etalchar{*}}24]{farhadzadeh2024neural}
\textsc{Farhadzadeh F., Hou Q., Le H., Said A., Rauwendaal R., Bourd A.,
  Porikli F.}:
\newblock Neural graphics texture compression supporting random access.
\newblock In \emph{European Conference on Computer Vision} (2024), Springer,
  pp.~412--429.

\bibitem[FNK94]{franti1994compression}
\textsc{Fr{\"a}nti P., Nevalainen O., Kaukoranta T.}:
\newblock Compression of digital images by block truncation coding: a survey.
\newblock \emph{The Computer Journal 37}, 4 (1994), 308--332.

\bibitem[Fow23]{fowler202extending}
\textsc{Fowler C.}:
\newblock Extending in-game textures using cdns for 'call of duty: Modern
  warfare 2'.
\newblock GDC'23, 2023.

\bibitem[HMB{\etalchar{*}}20]{hill2020physically}
\textsc{Hill S., McAuley S., Belcour L., Earl W., Harrysson N., Hillaire S.,
  Hoffman N., Kerley L., Patry J., Piek{\'e} R., et~al.}:
\newblock Physically based shading in theory and practice.
\newblock In \emph{ACM SIGGRAPH 2020 Courses}. 2020, pp.~1--12.

\bibitem[Jef24]{bolz2024cooperative}
\textsc{Jeff B.}:
\newblock Vk\_nv\_cooperative\_vector, 2024.
\newblock URL:
  \url{https://registry.khronos.org/vulkan/specs/latest/man/html/VK_NV_cooperative_vector.html}.

\bibitem[Khr25]{khronos2025image}
\textsc{Khronos}:
\newblock Vulkan documentation: Compressed image formats, 2025.
\newblock URL:
  \url{https://docs.vulkan.org/spec/latest/appendices/compressedtex.html}.

\bibitem[LdR14]{lagarde2015moving}
\textsc{Lagarde S., de~Rousiers C.}:
\newblock Moving frostbite to physically based rendering.
\newblock In \emph{SIGGRAPH 2014 Courses}. ACM, 2014.

\bibitem[NLP{\etalchar{*}}12]{nystad2012adaptive}
\textsc{Nystad J., Lassen A., Pomianowski A., Ellis S., Olson T.}:
\newblock Adaptive scalable texture compression.
\newblock In \emph{Proceedings of the Fourth ACM SIGGRAPH/Eurographics
  Conference on High-Performance Graphics} (2012), pp.~105--114.

\bibitem[PGM{\etalchar{*}}19]{paszke2019pytorch}
\textsc{Paszke A., Gross S., Massa F., Lerer A., Bradbury J., Chanan G.,
  Killeen T., Lin Z., Gimelshein N., Antiga L., Desmaison A., K\"{o}pf A., Yang
  E., DeVito Z., Raison M., Tejani A., Chilamkurthy S., Steiner B., Fang L.,
  Bai J., Chintala S.}:
\newblock \emph{PyTorch: an imperative style, high-performance deep learning
  library}.
\newblock Curran Associates Inc., Red Hook, NY, USA, 2019.

\bibitem[VSW{\etalchar{*}}23]{vaidyanathan2023random}
\textsc{Vaidyanathan K., Salvi M., Wronski B., Akenine-M{\"o}ller T., Ebelin
  P., Lefohn A.}:
\newblock Random-access neural compression of material textures.
\newblock \emph{arXiv preprint arXiv:2305.17105} (2023).

\bibitem[Wal92]{wallace1992jpeg}
\textsc{Wallace G.~K.}:
\newblock The jpeg still picture compression standard.
\newblock \emph{IEEE transactions on consumer electronics 38}, 1 (1992),
  xviii--xxxiv.

\bibitem[WDOHN24]{weinreich2024real}
\textsc{Weinreich C., De~Oliveira L., Houdard A., Nader G.}:
\newblock Real-time neural materials using block-compressed features.
\newblock In \emph{Computer Graphics Forum} (2024), vol.~43, Wiley Online
  Library, p.~e15013.

\bibitem[Z{\etalchar{*}}]{polyhaven}
\textsc{Zaal G., et~al.}:.
\newblock URL: \url{https://polyhaven.com/}.

\end{thebibliography}

%-------------------------------------------------------------------------
\newpage
% \appendix
% \section{Algorithms}

\definecolor{codegreen}{rgb}{0,0.6,0}
\definecolor{codegray}{rgb}{0.5,0.5,0.5}
\definecolor{codepurple}{rgb}{0.58,0,0.82}
\definecolor{backcolour}{rgb}{0.95,0.95,0.92}

\lstdefinestyle{myStyle}{
    language=C++,
    backgroundcolor=\color{backcolour},   
    commentstyle=\color{codegreen},
    keywordstyle=\color{magenta},
    numberstyle=\tiny\color{codegray},
    stringstyle=\color{codepurple},
    basicstyle=\ttfamily\footnotesize,
    breakatwhitespace=false,         
    breaklines=true,                 
    keepspaces=true,                 
    numbers=left,       
    numbersep=5pt,                  
    showspaces=false,                
    showstringspaces=false,
    showtabs=false,                  
    tabsize=2,
    frame=single,
    framexleftmargin=5mm, % Adjust left margin
    framextopmargin=3pt,  % Adjust top margin
}

\begin{algorithm*}
    \center
\begin{lstlisting}[language=C++, style=myStyle]
[numthreads(8, 4, 1)]
void ClassificationA(uint2 groupID: SV_GroupID, uint2 pixelCoords : SV_DispatchThreadID)
{   
    // Get the material ID of the pixel
    uint32_t matID = visibility_buffer_mat_id(pixelCoords)
    if (valid_mat_id(matID))
    {
        // First we need to find if there are multiple MLPs within this work group
        uint32_t maxID = WaveActiveMax(matID);
        uint32_t minID = WaveActiveMin(matID);

        // How many pixels valid within the WG?
        uint validPixelsCount = WaveActiveSum(1);

        // Flatten the work group ID
        uint wgID = uint(groupID.x + groupID.y * _TileSize.x);

        // Append in the right lists
        if (WaveIsFirstLane())
        {
            // This workgroup has only one material and at least 16 active pixels
            if (maxID == minID && validPixelsCount > 16)
                _UniformTileAppendBuffer.Append(wgID);
            else
                _ComplexTileAppendBuffer.Append(wgID);
            _ActiveTileAppendBuffer.Append(wgID);
        }
    }   
}

[numthreads(8, 4, 1)]
void ClassificationB(uint2 groupThreadID : SV_GroupThreadID)
{
    // Get the actual work group Index
    uint workGroupID = _ComplexTileConsumeBuffer.Consume();

    // Get it as coordinates
    uint wgX = workGroupID % _TileSize.x;
    uint wgY = workGroupID / _TileSize.x;

    // Compute the pixel coords
    uint2 pixelCoords = uint2(wgX * 8 + groupThreadID.x, wgY * 4 + groupThreadID.y);

    // Is this a valid pixel? If yes it needs to register
    uint32_t matID = visibility_buffer_mat_id(pixelCoords)
    if (valid_mat_id(matID))
    {
        // This is a mixed tile, so book a slot for this pixel in the right tile group
        uint targetSlot;
        InterlockedAdd(_MLPUsageBufferRW[matID], 1, targetSlot);

        // Get the the group offset
        uint32_t mlpGroupOffset = _MLPUsageBufferRW[UNIQUE_MLP_COUNT + matID];
        _IndexedTilesBufferRW[tileGroupOffset * WORK_GROUP_SIZE + targetSlot] = pixelCoords.x + pixelCoords.y * _ScreenSize.x;
    }
}
\end{lstlisting}
    \caption{
        \textbf{Tile-based classification.} \change{Our classification shader (here, in HLSL) use two passes: \texttt{classificationA} and \texttt{classificationB} to sort pixels into coherent tiles.}
        \label{alg:tile_based_classification}
    }
\end{algorithm*}

\end{document}